\documentclass[12pt]{iopart}

\usepackage{graphicx}
\usepackage{dcolumn}
\usepackage[final,dvips]{epsfig}
\usepackage{bm}
\usepackage{color}

\newcommand{\beq}{\begin{equation}}
\newcommand{\eeq}{\end{equation}}
\newcommand{\beqn}{\begin{eqnarray}}
\newcommand{\eeqn}{\end{eqnarray}}
\newcommand{\eqqref}[1]{Eq.~(\ref{#1})}

\newcommand{\lhb}{lower Hubbard band }
\newcommand{\uhb}{upper Hubbard band }

\newcommand{\om}{\ensuremath{\omega}}

\newcommand{\ttr}{{\rm Tr\,}}

\newcommand{\vv}[1]{{\mathbf #1}}
\newcommand{\up}{{\uparrow}}

\newcommand{\down}{\downarrow}

\newcommand{\g}{{\sigma}}

\newcommand{\ia}[1]{\left< #1 \right>}

\newcommand{\cd}{}

\newcommand{\pit}{\ensuremath{(\pi/2,\pi/2)}{}}
\newcommand{\pio}{\ensuremath{(\pi,0)}{}}

\newlength{\ccwidth}
\begin{document}

\title[Phase diagram and single-particle spectrum of CuO$_2$ layers]{ 
Phase diagram and single-particle spectrum of CuO$_2$ layers
within a variational cluster approach
to the 3-band Hubbard model.
}

\author{E. Arrigoni}

\address{
Institute of Theoretical Physics and Computational
Physics, Graz University of Technology, Petersgasse 16, 8010
Graz, Austria
}

\author{M. Aichhorn}

\address{
Centre de Physique Th\'{e}orique, \'{E}cole Polytechnique, 91128
Palaiseau Cedex, France 
}

\author{M. Daghofer}
\address{Department of Physics and Astronomy, The University of Tennessee, Knoxville, TN
 37996}
\address{Materials Science and Technology Division, Oak Ridge National Laboratory, Oak Ridge, TN 32831}

\author{W. Hanke}

\address{
Institute for Theoretical Physics, University of
W\"urzburg, Am Hubland, 97074~W\"urzburg, Germany
}

\begin{abstract}

We carry out a detailed numerical study of the
three-band Hubbard model in the underdoped region both in the
hole- as well as in the electron-doped case by means of the
variational cluster approach.
Both the phase diagram and the low-energy single-particle spectrum are
very similar to recent results for the single-band Hubbard
model with next-nearest-neighbor hoppings.
In particular, we obtain a mixed antiferromagnetic+superconducting
phase at low doping with a first-order transition to a pure
superconducting phase accompanied by phase separation.
In the single-particle spectrum 
a clear 
Zhang-Rice singlet band 
with an incoherent and a coherent part
can be seen,
in which
holes enter upon doping around \pit.
The latter
is very similar to the 
coherent quasi-particle band crossing the Fermi surface
in the single-band model.
Doped electrons go instead into the \uhb, first filling
the regions of the Brillouin zone around \pio.
This fact can  be related to the enhanced robustness of 
the antiferromagnetic phase as a function of electron doping compared
to hole doping.

\end{abstract} 
 

\maketitle

\section{Introduction}

The three-band Hubbard (3BH) Hamiltonian contains a minimal set of relevant
orbitals to
describe the physics of the copper-oxide layers in the high-temperature 
superconducting cuprates, namely the Copper $d_{x^2-y^2}$ and 
the oxygen $p_x$ and $p_y$ orbitals~\cite{emer.87}.
By eliminating the oxygen degrees of freedom, one can further
approximately reduce this effective Hamiltonian 
to the single-band Hubbard and to the $t-J$ model\cite{zh.ri.88}, which already
contain essential ingredients to describe a correlated system.
Having less degrees of freedom, these latter models can be more easily treated
via numerical methods such as Exact Diagonalisation (ED) or Quantum
Monte Carlo (QMC). Moreover, they already describe appropriately a number of
relevant physical properties of the 
 high-Tc cuprates (see, for ex., Refs. \cite{se.la.05,ai.ar.06,im.fu.98,wh.sc.99.cb,be.ca.00})
Nevertheless, because of the approximate reduction, 
it is not clear whether the 3BH model contains
important physics that is not described by a single-band Hubbard (1BH) Hamiltonian.
The nature of the insulating state at half filling is fundamentally different in the
two models in the relevant parameter range: while the  1BH model
describes a Mott insulator, in the 3BH model one has a charge-transfer
insulator.
More differences can be seen upon doping.
In a 3BH model there is a {\em qualitative} difference between hole
and electron doping: while doped holes go into the oxygen orbitals
forming the famous Zhang-Rice singlet~\cite{zh.ri.88}, doped 
electrons go into the Cu 
d$_{x^2-y^2}$ orbitals. 
This is believed to be the reason for the
better stability of the antiferromagnetic (AF) phase 
when doping with electrons with respect to doping with holes:
while introducing electrons in the Cu-orbitals merely produces a
dilution of the spins, whereas holes on oxygen sites produce a
ferromagnetic coupling between neighboring Cu orbitals, which is more
effective in destroying the AF order. 
While it is argued that one can incorporate this intrinsic
electron-hole asymmetry of the 3BH model into the effective 1BH model by
appropriate hopping terms~\cite{se.la.05,ai.ar.06},
a detailed
comparison between the 3BH model and such approximation 
is, so far, lacking.

Quite generally, the phase diagram of the high-temperature
superconductors (HTSC) displays a variety of competing phases at low
temperatures or energies, i.\,e., AF behavior, stripes, pseudogap
behavior and $d_{x^2-y^2}$ pairing. These phases are nearly degenerate
in energy. Therefore, in the theoretical modeling a change such as
the occurrence of $p-d$ charge fluctuations, included in the 3BH model
but not in the 1BH model, can tip the balance. Furthermore, the
balance can be tipped between the competing orders by using different
sets of parameters for a given model and/or by using different
techniques for solving this model. This is documented in numerical
studies already for the 1BH and t-J models, which have shown how
delicately balanced these models are between the nearly degenerate
phases~\cite{im.fu.98}. For example, altering the next-nearest-neighbor
hopping $t_{nnn}$ or the strength of the on-site correlation $U$ can favor $d_{x^2-y^2}$ pairing
correlations over stripes~\cite{wh.sc.99.cb}. Similarly, the delicate balance
is also reflected in the different results obtained using different
numerical techniques for the same model. For example, the cluster
lattice sizes and the boundary conditions may frustrate stripe
formation~\cite{be.ca.00}.
For the 3BH model early QMC studies have produced a set of model
parameters that consistently describe salient features of the HTSC,
such as the magnitude of the charge-transfer gap and the $T$- and
doping dependence of the normal-state magnetic
susceptibility~\cite{do.mu.90,do.mu.92}.
On the other hand, the well-known minus sign
problem embedded in the QMC calculations did not allow for simulations
searching, in particular, for
a superconducting (SC) state 
in the low-temperature limit for arbitrary fillings in the Hubbard
models.

Cluster techniques provide a controlled way to systematically approach
the infinite-size (and, thereby, low-energy) limit. Recently, progress
has been obtained  with the variational cluster approach (VCA), which
was proposed and used by Potthoff and our group~\cite{po.ai.03,da.ai.04}. This approach
provides a rather general and controlled way to go to the
infinite-sized lattice fermion system at low temperatures and at
$T=0$, in particular. The ground-state phase diagram of the
two-dimensional 1BH model was calculated within VCA by Senechal et
al.~\cite{se.la.05}, and by our
group~\cite{ai.ar.05,ai.ar.06}. 
There are technical
differences in the two works, but the overall results are
similar. For the cluster sizes used (up to 10 sites) in the VCA, the
$T=0$ phase diagram of the 1BH model correctly reproduces salient
features of the HTSC, such as the AF and $d$-wave SC ground state in
doping ranges, which are qualitatively in agreement with both
electron- and hole-doped cuprates. So it appears that much of the
difference between electron- and hole-doped cuprates, in particular, the
different stability of the AF phase, can be accounted for by a simple
1BH model, in which particle-hole symmetry is broken by a
next-nearest-neighbor hopping term, i.\,e. $t_{nnn}$.

A central question to be studied in the present paper is what is the
role of the oxygen degrees of freedom, of $p-d$ charge fluctuations
etc. on the $T=0$ phase diagram and the corresponding single-particle
excitations. If the results of this 3BH study are similar to the above
1BH model case, one is tempted to further argue in favor of the
dominant role played by spin fluctuations (which to a first
approximation are equally well captured in both models) over the role
of charge fluctuations. Up to what doping does this hold?

Correspondingly, 
in this paper, we carry out a detailed calculation of the phase
diagram and of the single-particle spectrum of the 3BH model from the
hole- to the electron-doped region. We use the above recently proposed 
cluster variational method which is appropriate to treat
strongly-correlated systems, i.\,e., the
VCA~\cite{po.ai.03,da.ai.04}.
We allow both for a AF  and for a SC phase, as well as for a mixed
AF+SC one, as obtained for the 1BH model~\cite{se.la.05,ai.ar.05}.
In the different doping range from underdoped to optimally and overdoped 
in both electron and hole-doped regions 
we evaluate the single-particle spectrum.
The main outcome of our calculations is 
that indeed the physics of the 3BH and of the 1BH
model is very similar concerning both the phase diagram as well as the
single-particle spectrum.
More specifically, at low doping we obtain a mixed AF+SC phase
in which the transition to the pure SC phase is first order as a
function of chemical potential $\mu_h$, and, therefore, accompanied by
phase separation. The critical doping at which the AF+SC phase disappears
is qualitatively consistent with experimental results being larger
in the electron-doped case.
Also the single-particle spectrum in the vicinity of the gap 
looks very similar to the one
obtained for the 1BH model, whereby the \lhb
is replaced by the dispersive Zhang-Rice band.

Related recent works address the asymmetry between electron- and
hole-doping in the three-band Hubbard model without oxygen-oxygen interaction:
In Ref.~\cite{de.wa.08u}  an asymmetry between the quasiparticle mass enhancement
in the electron and hole-doped region was found, which however gets compensated 
in the infrared optical spectral weight.
In Ref.~\cite{we.ha.08u}, the asymmetry between electron and hole
doping, and the presence of waterfall structures was discussed within
a Local Density Approximation (LDA)+Dynamical Mean Field Theory
approach to the three-band Hubbard model.
Ref.~\cite{ke.sa.08u} contains a combined local
density-functional theory and DCA (dynamical cluster approach) for
3BH-models for hole-doped cupraes. One of the important outcomes of
this study is that the occurrence of SC transition depends rather
sensitively on the ``down-folding'', i.e. whether the orbital basis
set is more or less localized (with the latter reproducing the LDA
bands over a larger energy window). A very relevant paper, similar
in spirit albeit at finite temperature, is the work by Macridin et.
al. \cite{ma.ja.05.pc}, addressing the low-energy physics of cuprates within
a two-band Hubbard model and again checking for the validity of the
one-band Hubbard model. Here, only oxygen states that hybridize
directly with Cu are considered. Nevertheless, the conclusion is
similar, in that a $t-t'-U$ single-band Hubbard model captures the
basic low energy (below $\sim 0.5 eV$) physics.
 For further work on the 3BH model see also, e.\,g., 
Refs.~\cite{ho.st.89,sc.sc.91,un.fu.93,lu.bi.93,lo.av.96,wa.do.98,th.gr.07u,de.wa.08u,we.ha.08u,ke.sa.08u}.

Our paper is organized as follows:
In Sec.~\ref{model} the model we are using is presented and  the VCA
method summarized. 
Results for the phase diagram and for the
single-particle spectrum 
are shown 
in Sec.~\ref{results}.
Finally, in Sec.~\ref{summary} we draw our conclusions.

\section{Model and method}
\label{model}

It is commonly believed that the relevant physics of high-Tc
superconductors takes place mainly on the copper-oxide layers. 
A minimal model for these layers contains the $d_{x^2-y^2}$ orbital as well
as a $p_x$ and $p_y$ orbital per unit cell.
The electron dynamics in these orbitals is described by the 
three-band (Emery) model, which, in hole notation,  reads
 \cite{emer.87}

\begin{figure}[h]
  \centerline{
%
\includegraphics[width=0.7\ccwidth]{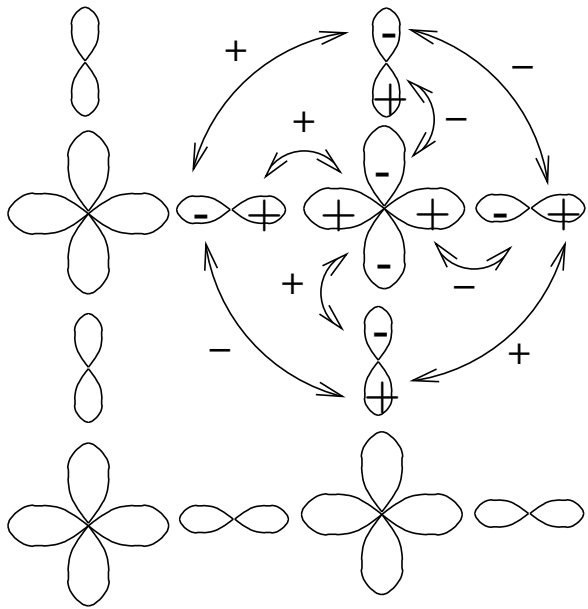}
}
  \caption{\label{cuo2p}
    Orbitals in the three-band Hubbard model with phase convention.}
\end{figure}

\beqn
&&  H_{3b}=
(\epsilon_d-\mu_h)
\sum_{R \g} n^d_{R \g} 
\nonumber\\ && + 
t_{pd}\sum_{\ia{R S} \g} 
  \alpha_{R S} ( p^\dagger_{S \g} 
    d_{R\g}  + H.C.
 )
\nonumber\\ &&
+  t_{pp}\sum_{\ia{\ia{SS'}} \g} \alpha'_{SS'} 
  ( p^\dagger_{S \g} p_{S'\g} + H.C.
)
\nonumber\\ &&
  + 
(\epsilon_p-\mu_h)
\sum_{S\g}  p^\dagger_{S\g}
  p_{S\g} 
  +U_{dd}\sum_{R} \cd n^{d}_{R \up} n^{d}_{R \down} +
\nonumber\\&& 
+ U_{pd}\sum_{\ia{R S} \g \g'} 
 n^{p}_{S\g}  n^{d}_{R \g'} 
\nonumber\\&& 
+ U_{pp}\sum_{S} 
 n^{p}_{S\up}  n^{p}_{S \down} \;. 
\label{3bh}
\eeqn
Here, $d^\dagger_{R\g}$ and $p^\dagger_{S\g}$ create a
{\em hole}
in
the copper $3d_{x^2-y^2}$ and in the oxygen $2p_{\delta}$ orbital 
($\delta=x,y$ depending on the position with respect to the Cu site),
respectively, and
$ n^{d}_{R \g}$, $n^{p}_{S\g}$ are the corresponding occupations.
 $R$ denote copper and $S$ oxygen lattice sites.
$\ia{R S}$ limits the sum over
next-neighbor Cu-O  and
$\ia{\ia{S S'}}$ over next-neighbor O-O lattice sites. 
$\epsilon_d$ and $\epsilon_p$
are the on-site energies of the copper and the oxygen orbitals,
respectively, and $\mu_h$ is the hole chemical potential. 
$t_{pd}$ is the Cu-O and $t_{pp}$ is the direct O-O hopping amplitude.
We set the zero of the energy such that we have $\epsilon_d=0$ on the copper site,
and introduce  the charge-transfer gap $\Delta=\epsilon_p-\epsilon_d$.
The Coulomb
repulsion between two holes is taken into account by the 
terms
 $U_{dd}$, $U_{pp}$, and $U_{pd}$, when the two holes sit on the same Cu orbital,
 on the same Oxygen orbital, or on two neighboring Cu and O orbitals, respectively.
$\alpha_{RS\delta}$ and $\alpha'_{SS'}$ describe the sign of the phases of the
Cu-O and O-O hopping due to the $d$ and $p$ symmetries of the Cu and O
orbitals, respectively (according to the convention in Fig.~\ref{cuo2p}).
In our calculation, we take 
 typical values of the parameters, as obtained by constrained
density-functional calculations~\cite{hy.sc.89}, and consistent with
earlier cluster calculations~\cite{mila.88},
as well as extensive Quantum-Monte-Carlo studies~\cite{do.mu.90,do.mu.92}.
In units of the Cu-O hopping $t_{pd}\equiv t$, we thus have
$t_{pp}=0.5$, $\Delta=3.0$,  $U_{dd}=8.0$, $U_{pd}=0.5$, and
$U_{pp}=3.0$. 

Numerically exact methods such as Quantum Monte Carlo or exact
diagonalisation are restricted to relatively high temperatures or
small cluster sizes~\cite{ho.st.89}. 
The self-energy-functional approach (SFA)~\cite{pott.03}
provides a variational scheme
to use the information from the solution of an exactly solvable ``reference system''
(for example a small cluster) to 
approximate the properties of an infinite lattice.
With the help of this reference system, which must have the 
same interaction as the original one
(the infinite lattice), one can evaluate the grand potential 
$\Omega$
of the original system exactly within a restricted set of
self-energies.
The ``best'' solution thus is given by finding the saddle point of $\Omega$
within this set of self-energies.
Within the variational-cluster approach~\cite{po.ai.03,da.ai.04}, the reference system 
is obtained by splitting the original lattice into small clusters
and by adding single-particle $\vv t'$ terms which parametrize the variation of
the self-energy.
The SFA potential $\Omega$ of the original system can be evaluated exactly for
the self energies $\Sigma$ accessible to the reference systems 
and is given by
\beq
\label{omega}
\Omega = \Omega' + \ttr \ln (\vv G_0^{-1} -\vv \Sigma)^{-1} - \ttr \ln \vv G'
\eeq
Here, $\vv G_0$ is the free Green's function of the model given by \eqqref{3bh},
$\Omega^\prime$, $\vv \Sigma$, and $\vv G^\prime$ are the grand
canonical potential, 
the self-energy and the Green's function of the cluster reference system
which depends on the 
 the single-particle parameters $\vv t'$.
In the present study we 
consider the 12-sites cluster (i.\,e., a $2\times 2$ unit cell) shown in Fig.~\ref{cuo2p}
as a reference system and
 search for the
stationary solution characterized by the condition $\partial \Omega / \partial 
\vv t' = 0$.
This stationary point provides a good approximation to the exact
solution for the system in the thermodynamical 
limit provided the self-energy is sufficiently ``short ranged'',
i.\,e., sufficiently localized within the cluster. In the limit in
which the self-energy is
exactly localized within the cluster, the VCA becomes an exact solution.

Since we want to describe the transition from the AF phase at low
doping to the SC phase at higher doping, the reference system used in
the VCA consists
of the  $2\times2$ CuO$_2$ unit-cell, displayed in Fig.~\ref{cuo2p}. It is
described by the Hamiltonian~\eqqref{3bh} plus the additional
``Weiss-field'' terms
\beqn
\Delta H &=& h_{AF} \sum_{R, \g} \g \eta^{AF}_R
\ n^d_{R \g} 
\nonumber\\ &&
+ \frac{h_{SC}}{2} \bigl[
\sum_{R r} 
(\g \eta^{d}_r \ d_{R,\g} d_{R+r,-\g} + H.C.)
\nonumber\\  &&
+
\sum_{S r} 
(\g \eta^{d}_r p_{S,\g} p_{S+r,-\g} +H.C.)
\bigr]
\nonumber\\  &&
\label{delh}
+
\Delta \epsilon \left[
\sum_{R \g} n^d_{R \g} + \sum_{S \g} n^p_{S \g} \right] \;.
\eeqn
Here,  $r=\pm \vv e_x, \pm \vv e_y$ (we take the size of the unit cell
$a=1$), and the d-wave phase factor  
$\eta^{d}_r=+1$ for 
$r =\pm \vv e_x$, and 
$\eta^{d}_r=-1$ for 
$r =\pm \vv e_y$.  
The AF order is induced by $\eta^{AF}_R= \exp( i \vv Q \cdot \vv R) $ with $\vv Q = (\pi,\pi)$.
The fields controlling the variation of
the self-energy in the reference system are the AF 
 ($h_{AF}$) and the  the d-wave SC  ($h_{SC}$) ``Weiss''
fields~\cite{se.la.05,ai.ar.05}, as well as 
a shift in the on-site energy 
($\Delta \epsilon$), which is required in order to describe
consistently the particle density~\cite{ai.ar.06}.
We stress that these Weiss fields are only present in the reference
system and are used in order to optimize the self-energy to be used in the
original lattice model.
In principle, one could introduce additional variational parameters
such as a different on-site energy shift and a different SC Weiss
field
 for $d$ and $p$ orbitals. 
However, too many variational parameters make the
numerical search for the saddle point of $\Omega$ too difficult and
time consuming. 

Notice that the
intercluster interaction terms cannot be treated exactly by VCA, 
so that-in principle-we are not able to treat $U_{pd}$ interactions beyond the
cluster of Fig.~\ref{cuo2p}. However, to neglect them completely would
be a quite rough approximation, since, for example, O orbitals at the
cluster boundary 
would be influenced by $U_{pd}$ much less than the ones in
 the center. In order to overcome this problem, we treat nonlocal
 interactions by periodic boundary conditions. This turned out to be
 quite accurate, whenever the system is not close to a charge-density
 wave phase transition~\cite{ai.ev.04}.

\section{Results}
\label{results}

We first determine the zero-temperature ($T=0$) 
phase diagram in the hole- and electron-doped
region by determining the optimal values of the variational parameters
$h_{AF}$, $h_{SC}$ and $\Delta \epsilon$  as a function of the chemical
potential $\mu_h$.
In general, there is always more than one solution for a given $\mu_h$:
we adopt the usual criterion of taking the solution with the lowest
grand-canonical energy $\Omega$.

\begin{figure}[h]
\centerline{\includegraphics[width=0.99\ccwidth]{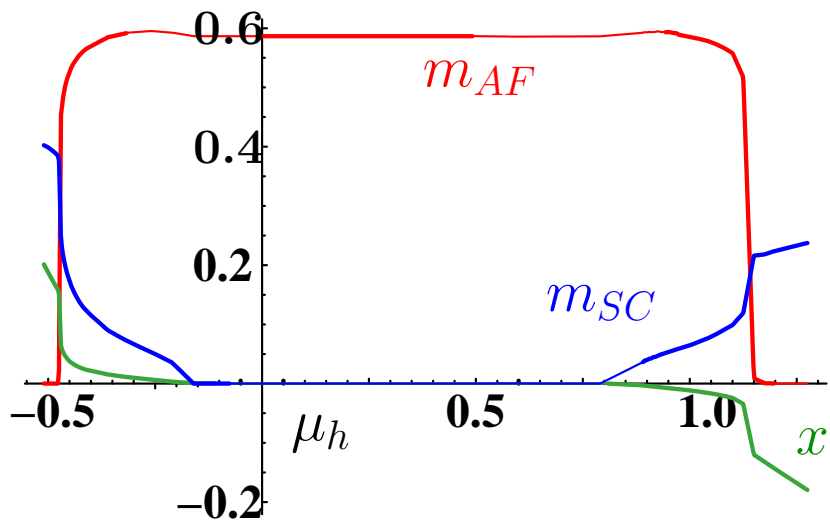}}
  \caption{\label{phase}  (Color online)
    AF ($m_{AF}$) and SC ($m_{SC}$) order parameters,
    and electron doping $x$ as a function of
  hole  chemical potential $\mu_h$.}
\end{figure}

\begin{figure}[h]
\centerline{\includegraphics[width=0.9\ccwidth]{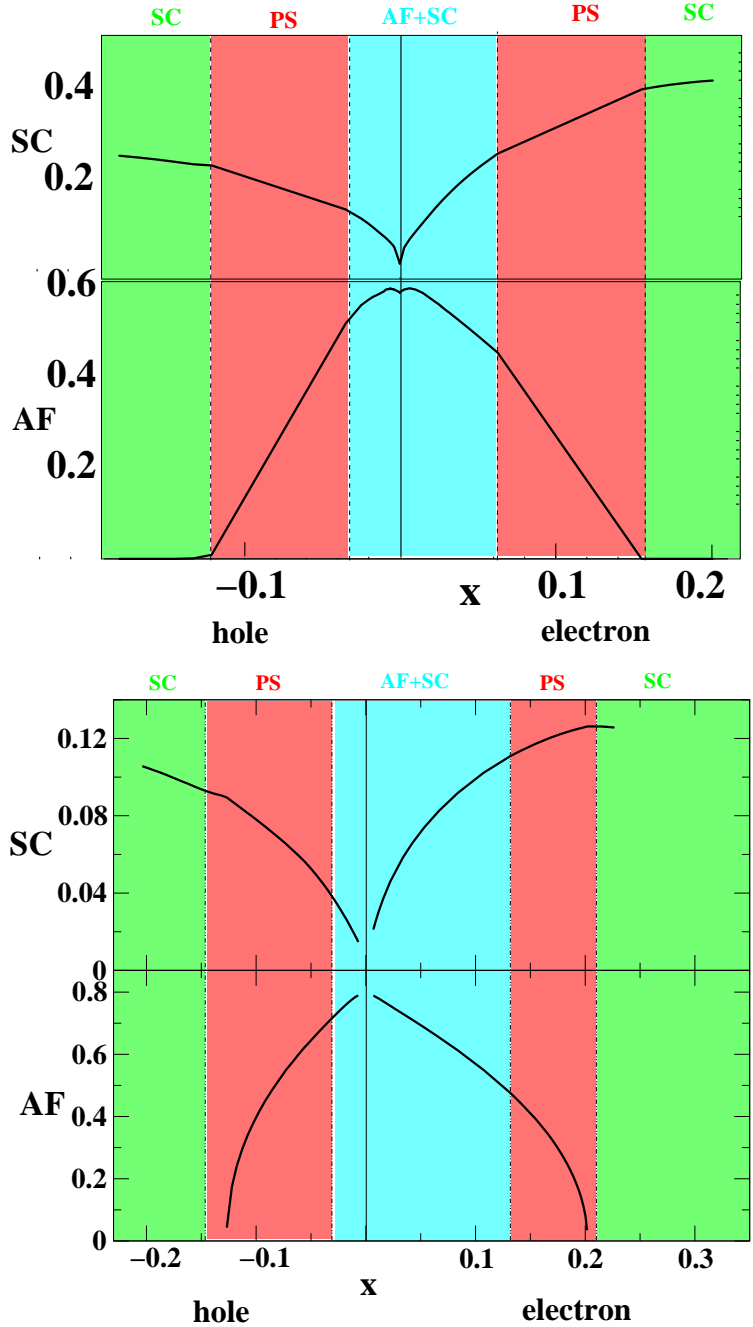}}
  \caption{\label{phasevsx}  (Color online)
    AF ($m_{AF}$) and SC ($m_{SC}$) order parameters 
    plotted as a function of
    electron doping $x$ for the three-band (upper panel) and
    single-band (lower panel) Hubbard model with comparable $2\times2$
    unit cells used as a reference system. 
    In the upper panel, values 
    in the phase-separation region, i.\,e., between the vertical dashed lines 
    marked by PS, have been
    linearly interpolated}
\end{figure}

In Fig.~\ref{phase} we plot the staggered magnetization (red color)
\beq 
m_{AF}=
\sum_{R \g} \g \eta^{AF}_R
<n^d_{R \g}>\;,
\eeq 
the d-wave superconducting order parameter (blue)
\beqn
m_{SC}&&=  \frac{1}{2} \bigl<
\sum_{R r} 
(\g \eta^{d}_r d_{R,\g} d_{R+r,-\g} + H.C.)
\nonumber \\&&
+
\sum_{S r} 
(\g \eta^{d}_r p_{S,\g} p_{S+r,-\g} +H.C.)
\bigr>
\eeqn
and the electron doping $x$  (green) 
as a function of
   chemical potential $\mu_h$.
Additionally, the order parameters are shown as function 
of doping $x$ in Fig.~\ref{phasevsx}.

The phase diagram is qualitatively and, in part, quantitatively very
similar to the one obtained for the single-band Hubbard
model~\cite{ai.ar.05,ai.ar.06,ai.ar.07.ps,ai.ar.07}. 
There is a mixed AF+SC phase at low doping both in the hole- as well
as in the electron-doped case. The AF phase ends at  a first order transition as a
function of chemical potential, which is accompanied by phase
separation.
In Ref.~\cite{ai.ar.06.vc} it was discussed that this phase separation
is likely to be persistent for the 1BH 
in the hole-doped case, while it seems to
disappear when considering larger clusters in the electron-doped case.
For the 3BH studied here, 
the critical doping at which the AF phase is destroyed is $3.5 \%$ in
the hole-doped and $6.5\%$ in the electron-doped case.
This is again in qualitative agreement with the experimental situation, in that
the AF phase is more stable for electron doping.
It is remarkable to observe such similarities between the 1BH and the
3BH model (cf. Refs.~\cite{se.la.05,ai.ar.05}) despite of the fact that
doping, and -- in particular -- the asymmetry between electron- and hole-doping is
fundamentally different in the two models.

\begin{figure*}
\begin{center}
\includegraphics[width=0.7\ccwidth]{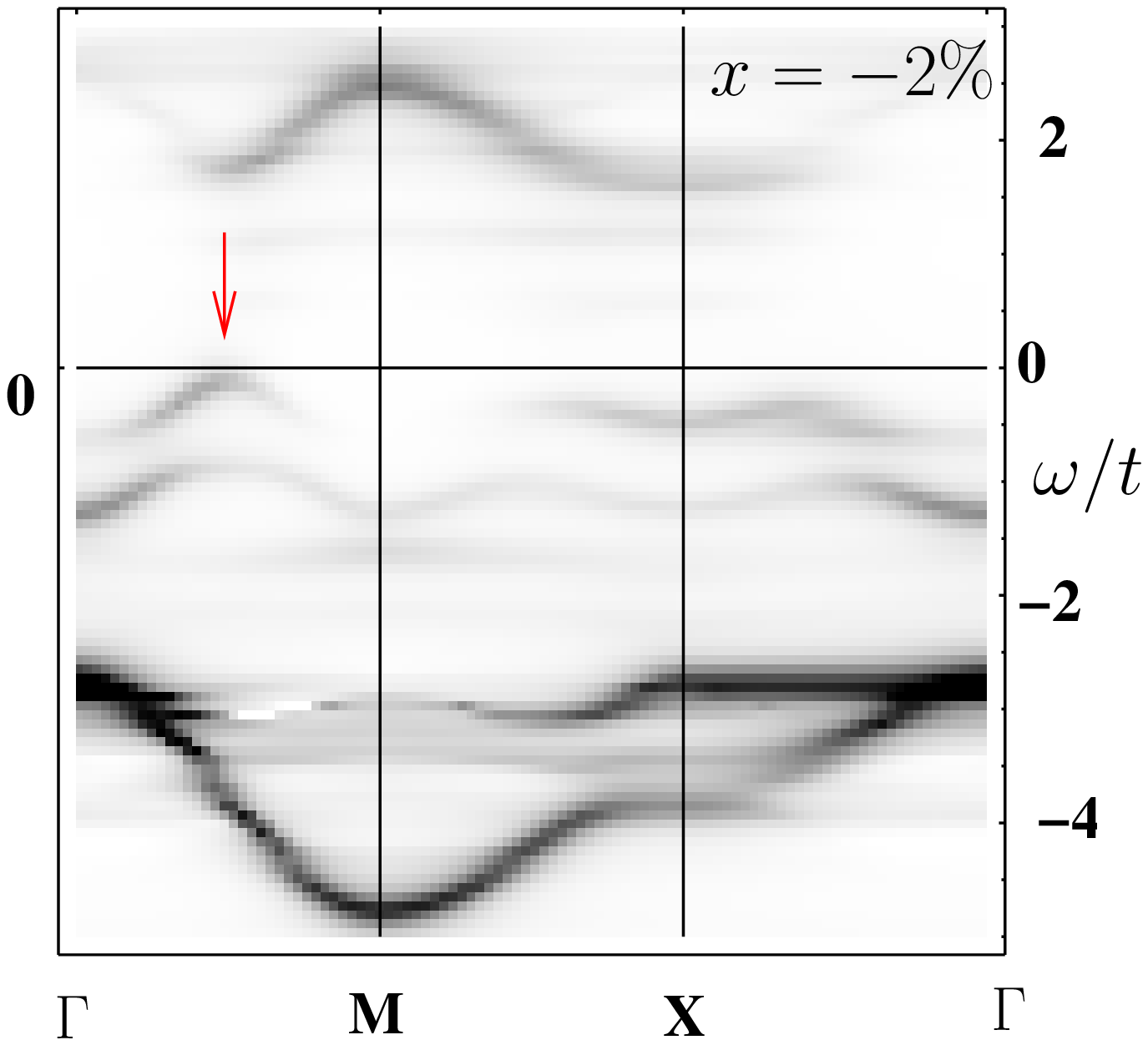} 
\includegraphics[width=0.7\ccwidth]{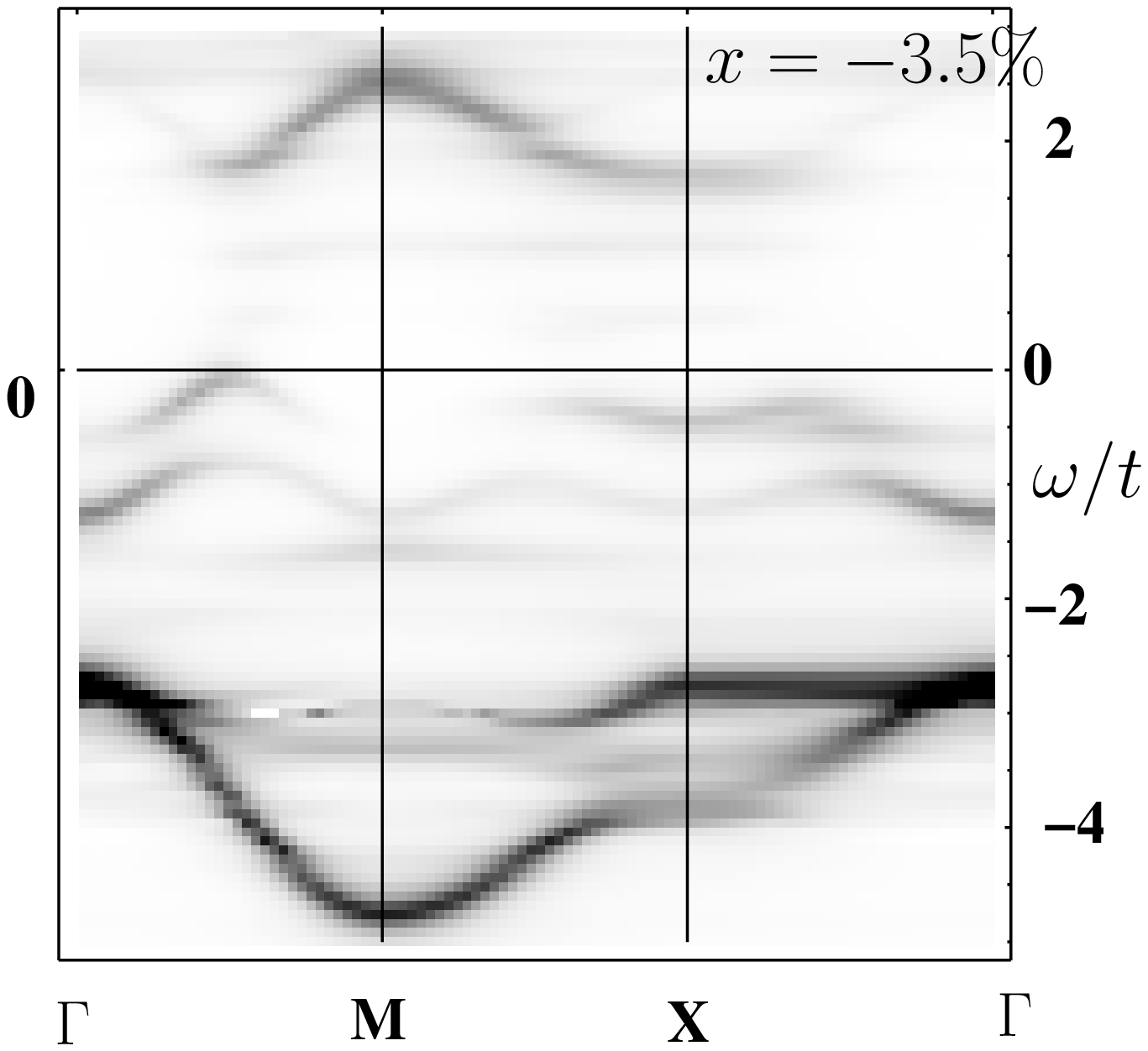} \\[0.05cm]
\includegraphics[width=0.7\ccwidth]{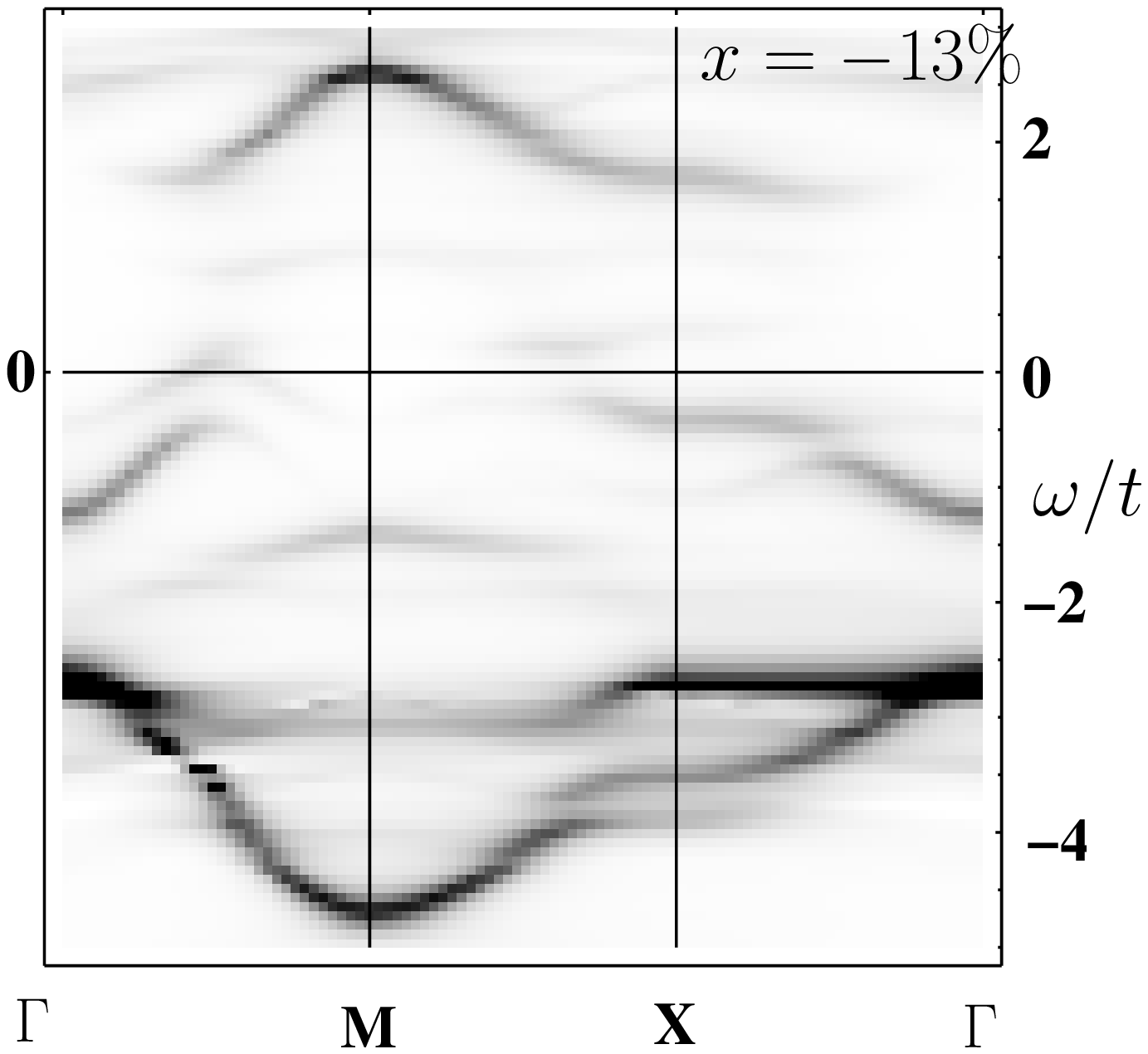} 
\includegraphics[width=0.7\ccwidth]{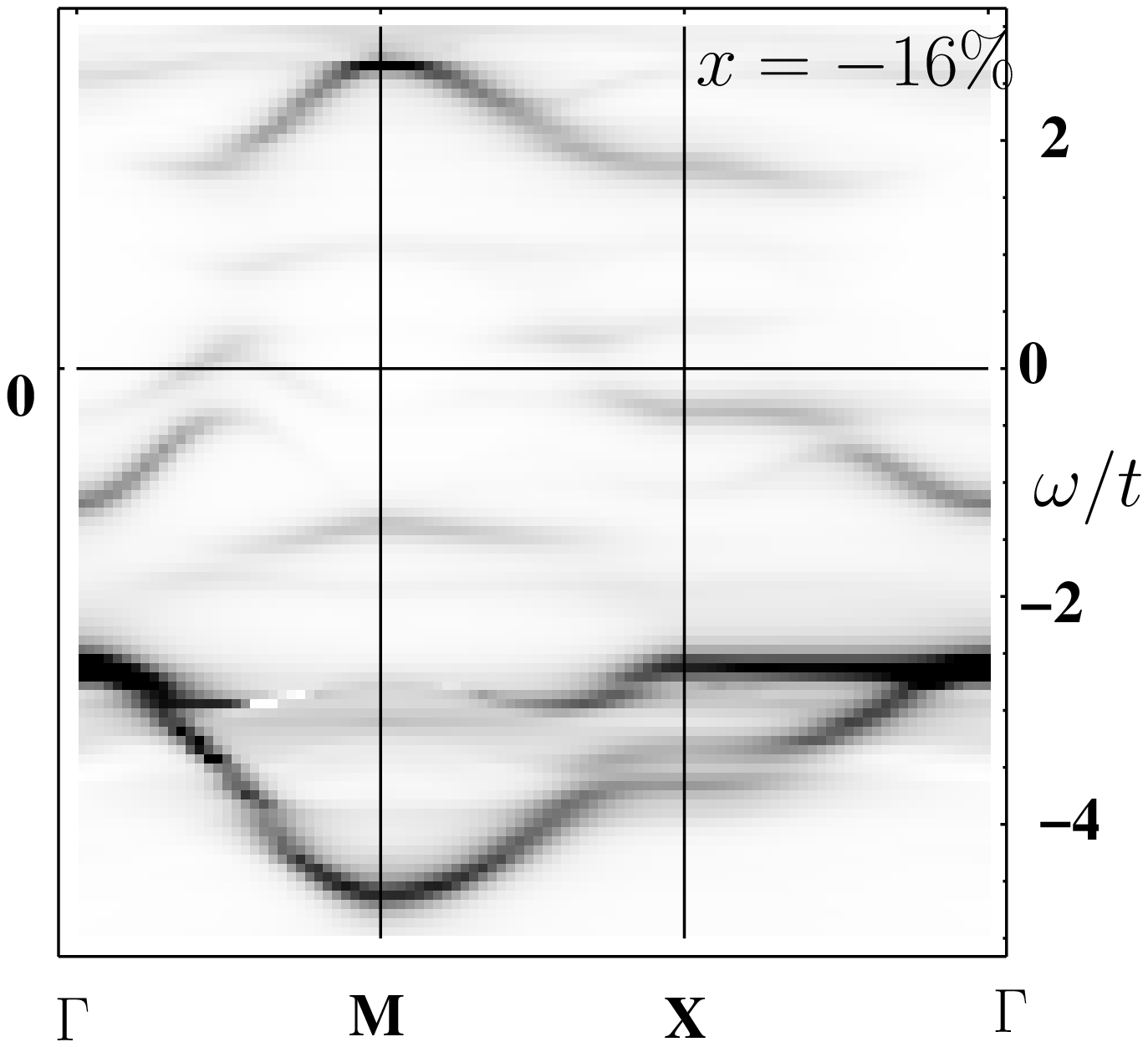}
\end{center}
\caption{\label{spectrh} (color online)
Evolution of the single-electron spectrum of the 3BH model for hole doping.
 The arrow for $x=-2\%$
 indicates where holes first enter upon doping. 
} 
\end{figure*}

\begin{figure*}
\begin{center}
\includegraphics[width=0.7\ccwidth]{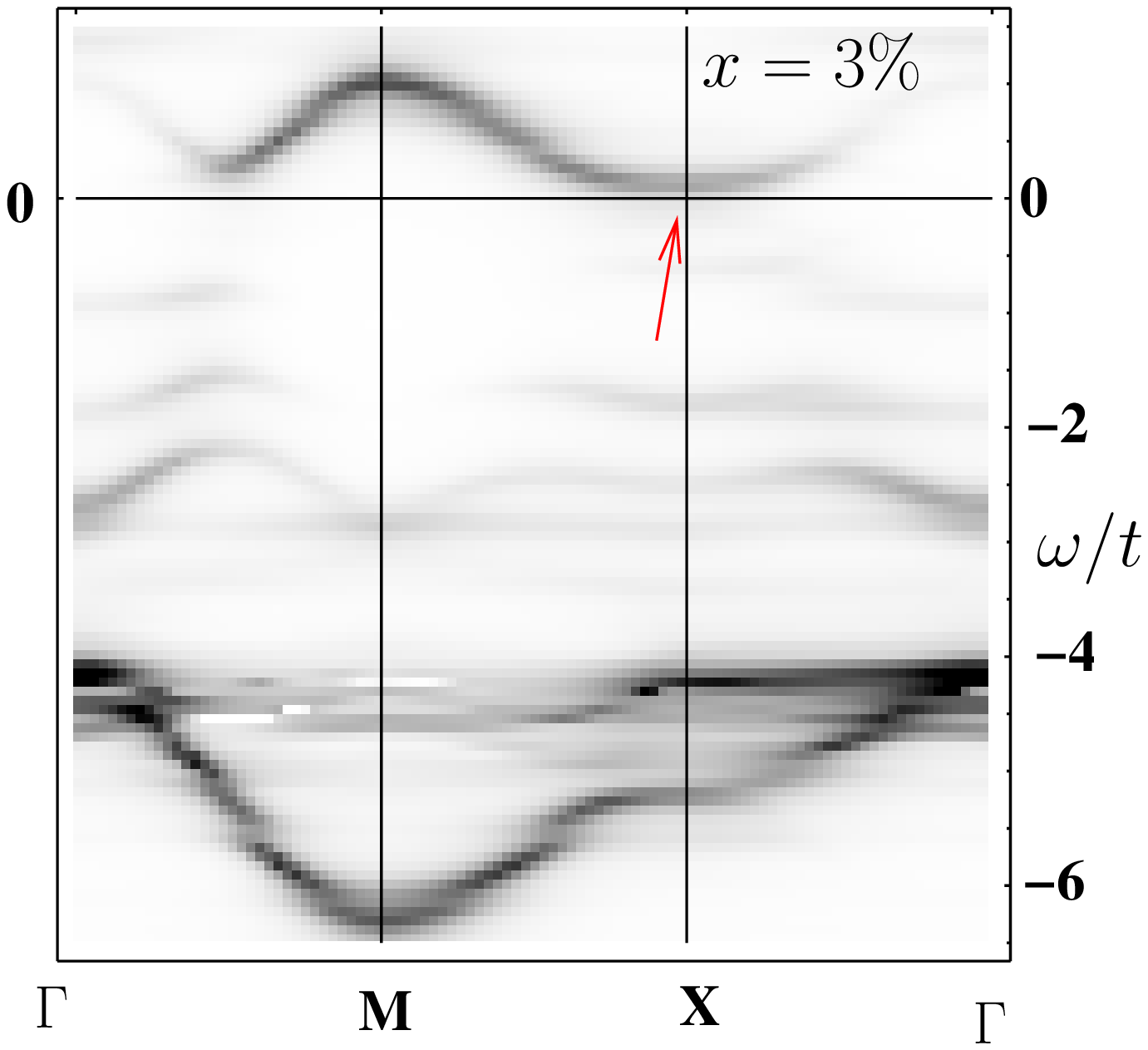} 
\includegraphics[width=0.7\ccwidth]{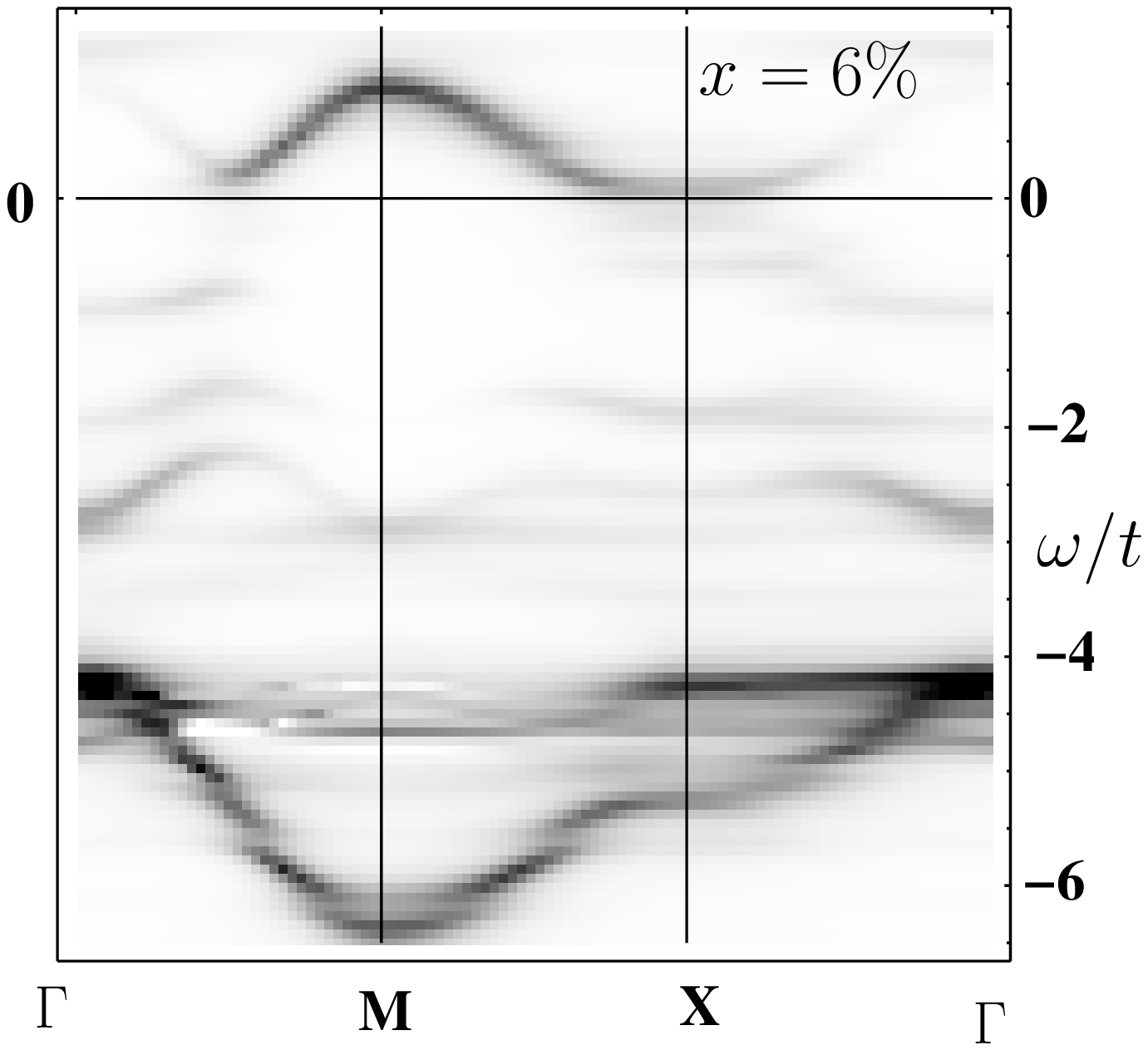} 
\includegraphics[width=0.7\ccwidth]{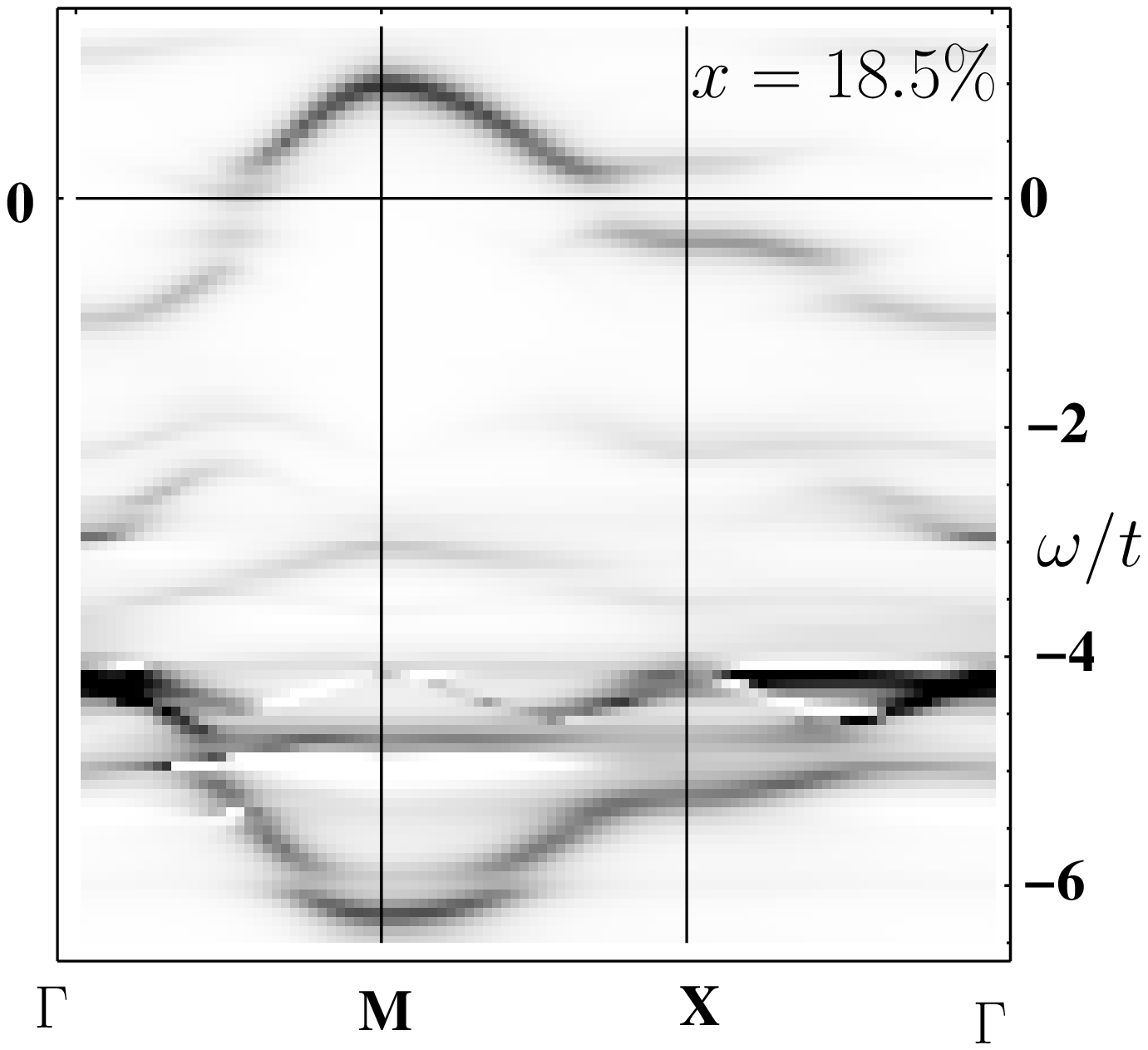}
\end{center}
\caption{
\label{spectre} (color online)
Evolution of the single-particle spectrum 
of the 3BH model
for electron doping. 
 The arrow for $x=3\%$
 indicates where electrons first enter upon doping.
} 
\end{figure*}

In order to explore the relation between single-particle excitations, their doping
evolution, and the phase
diagram of Figs.~\ref{phase} and ~\ref{phasevsx}, we plot in Figs.~\ref{spectrh} and
~\ref{spectre} the 
spectral function $A(\vec{k}, \omega)$ 
 of the 3BH model (in a grayscale plot) 
 for different dopings in
both the hole- (Fig.~\ref{spectrh}) and electron-doped
(~\ref{spectre}) case.

The spectrum in Fig.~\ref{spectrh}
shows the well-known features of the 3BH model. 
Starting from the top of the spectrum, we first find the 
\uhb 
 at $\om/t \approx 2.0$. 
Around the Fermi energy we can see
the Zhang-Rice singlet (ZRS) band~\cite{zh.ri.88}, with a total width
of about
$1$ (in units of $t$, see also Ref.~\cite{je.es.92}).
This band describes a dispersive singlet state 
 (ZRS) formed between a hole on the plaquette of the 
 O orbitals and one on the Cu site. It consists of a coherent part
 crossing the Fermi energy and of an incoherent part at higher binding
 energies. The separation between these two structures 
can be
 associated with the so-called ``waterfall'' structure observed in
 several high-Tc
 compounds~\cite{gr.gw.06u,ko.bo.07u,in.ko.07u,in.fi.07u}, see also
 Ref.~\cite{we.ha.08u}. 
Between the \uhb  and the ZRS there is an optical gap of about $1.5 t$,
which is consistent with the experimental value~\cite{ro.nu.90}.
At higher binding energies $\om/t < -3$ we find the two oxygen bands: a weakly
dispersive ``almost-non-bonding'' band (the weak dispersion is due to
$t_{pp}$), and the dispersive antibonding band. The 
lower Hubbard band
is located approximately at the same binding energy.

\begin{figure*}
\begin{center}
\includegraphics[width=1.4\ccwidth]{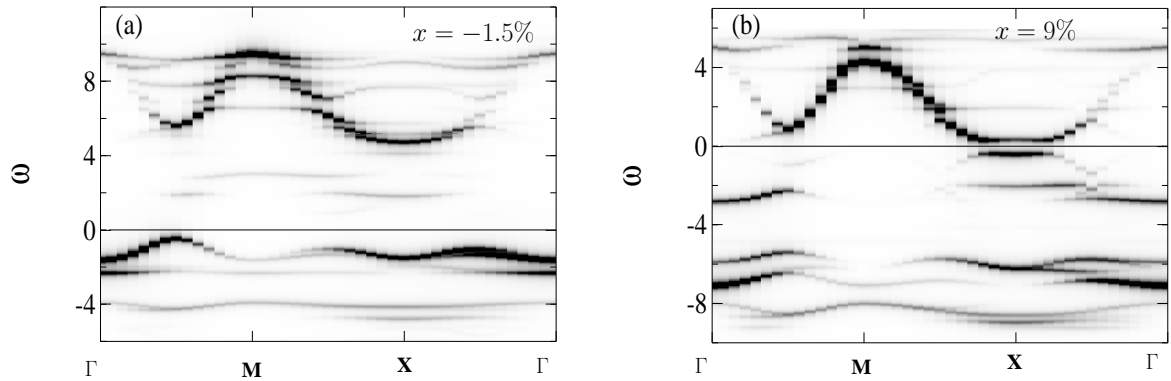} 
\end{center}
\caption{
\label{1band}
Single-particle spectrum for the 1BH model for hole
(a) and electron (b) doping (adapted from Ref.~\cite{ai.ar.06}). 
} 
\end{figure*}

At low energies,
the similarity with the spectrum of the 1BH model is remarkable, as
can be seen by comparing figures \ref{spectrh} and \ref{spectre}
with the spectrum of the 1BH model plotted in Fig.~\ref{1band} for comparison
(see also Refs.~\cite{se.la.05,ai.ar.05,ai.ar.06}). 
Notice that when comparing the band structure between the 1BH and the
3BH models,
the ZRS band takes here
the role of the 
coherent quasi-particle band in the 1BH model.
On the other hand, the \uhb of the 1BH model is replaced
with the \uhb in the 3BH model.
In the hole-doped case (Fig.~\ref{spectrh}), holes first enter the ZRS
band around \pit, and form hole pockets
at low doping.
Here, metallic screening from nodal electrons makes the AF solution quickly unstable.
We find a different situation for electron doping: The additional
electrons are first doped 
around \pio in the \uhb (Fig.~\ref{spectre}). Here, the AF solution
remains more stable,
since there is a SC gap
with maximal size at the anti-nodal point. Thus the screening is
less effective, until - as a function of doping 
electrons start to reach the \pit\ point.
Again these results are consistent wit 
the 1BH model,  and also with experiments
(for a review see Ref. \cite{da.hu.03}).

\begin{figure*}
\begin{center}
  \includegraphics[width=0.7\ccwidth]{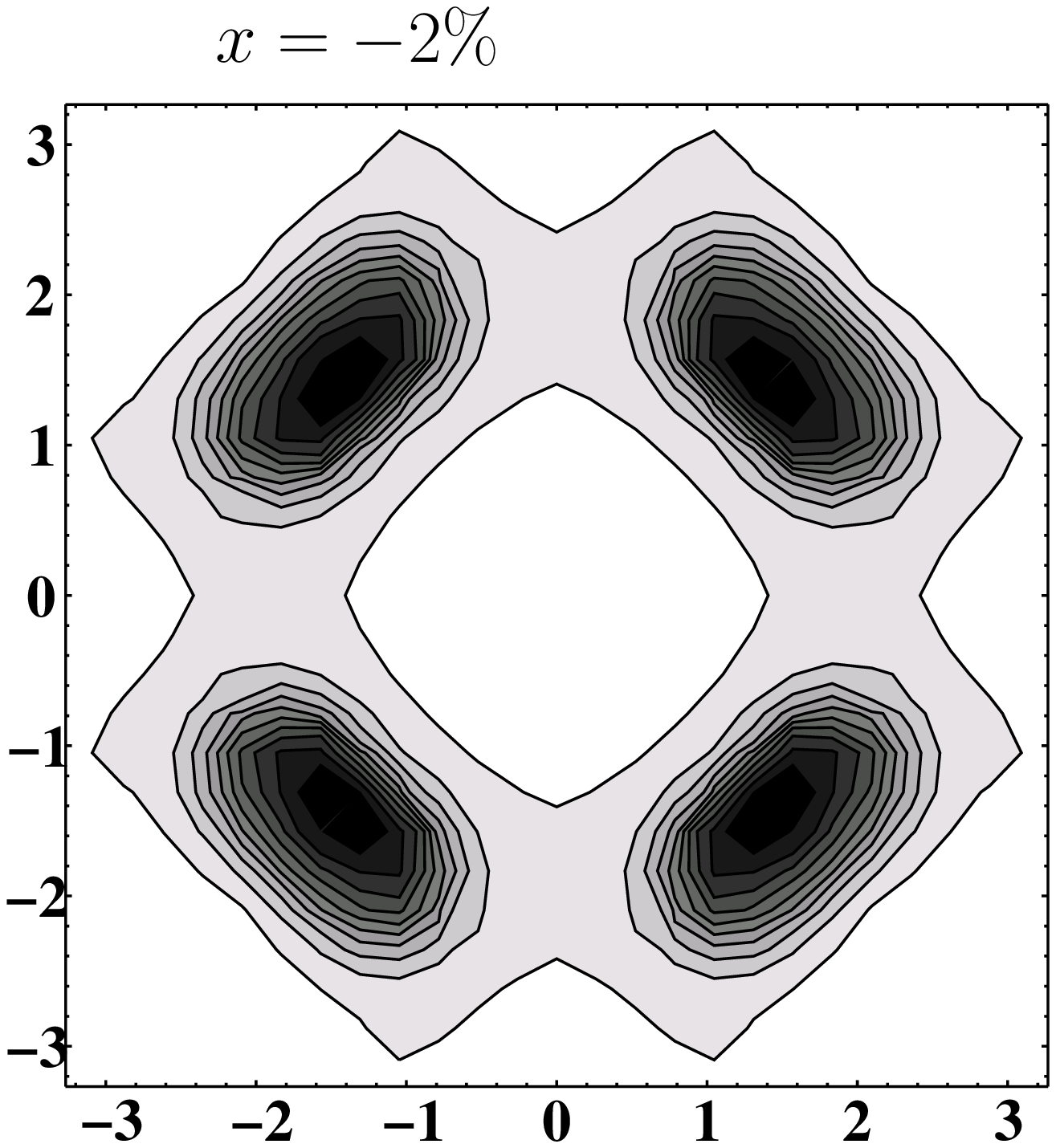} 
  \includegraphics[width=0.7\ccwidth]{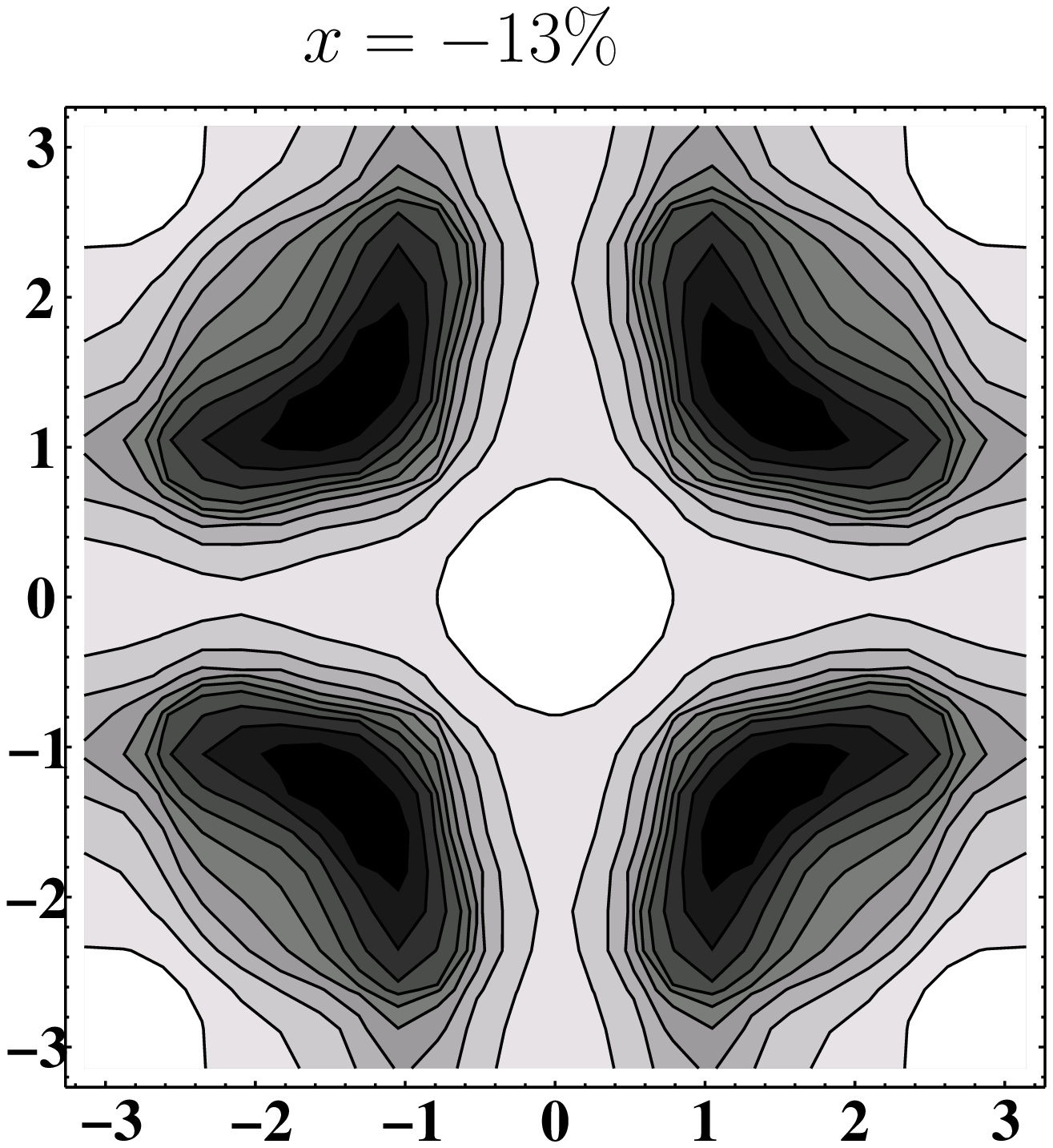} 
\end{center}
\caption{\label{fsh}
Low-energy integrated spectrum of the 3BH model for two different hole dopings.
}
\end{figure*}

\begin{figure*}
\begin{center}
  \includegraphics[width=0.7\ccwidth]{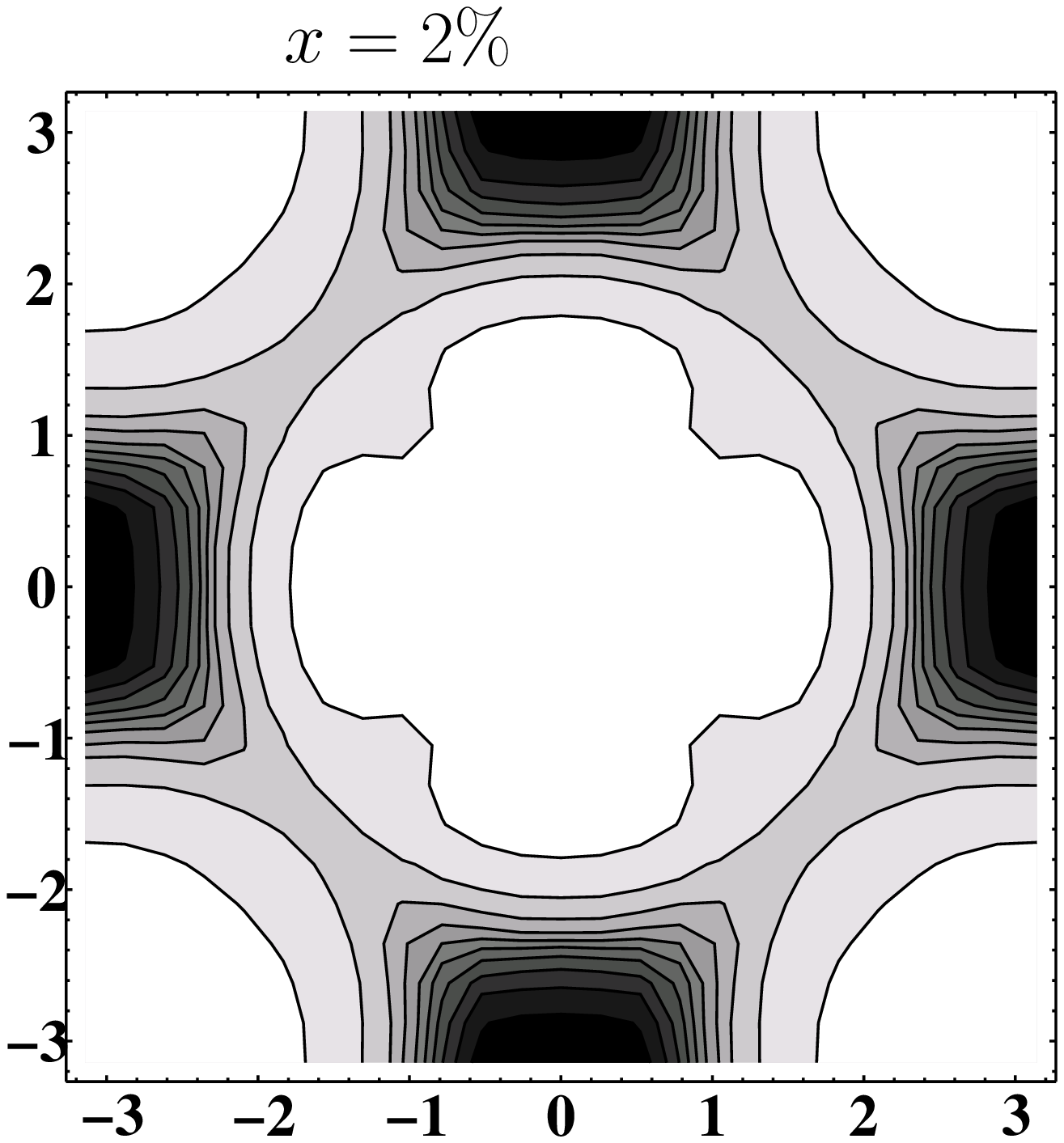} 
  \includegraphics[width=0.7\ccwidth]{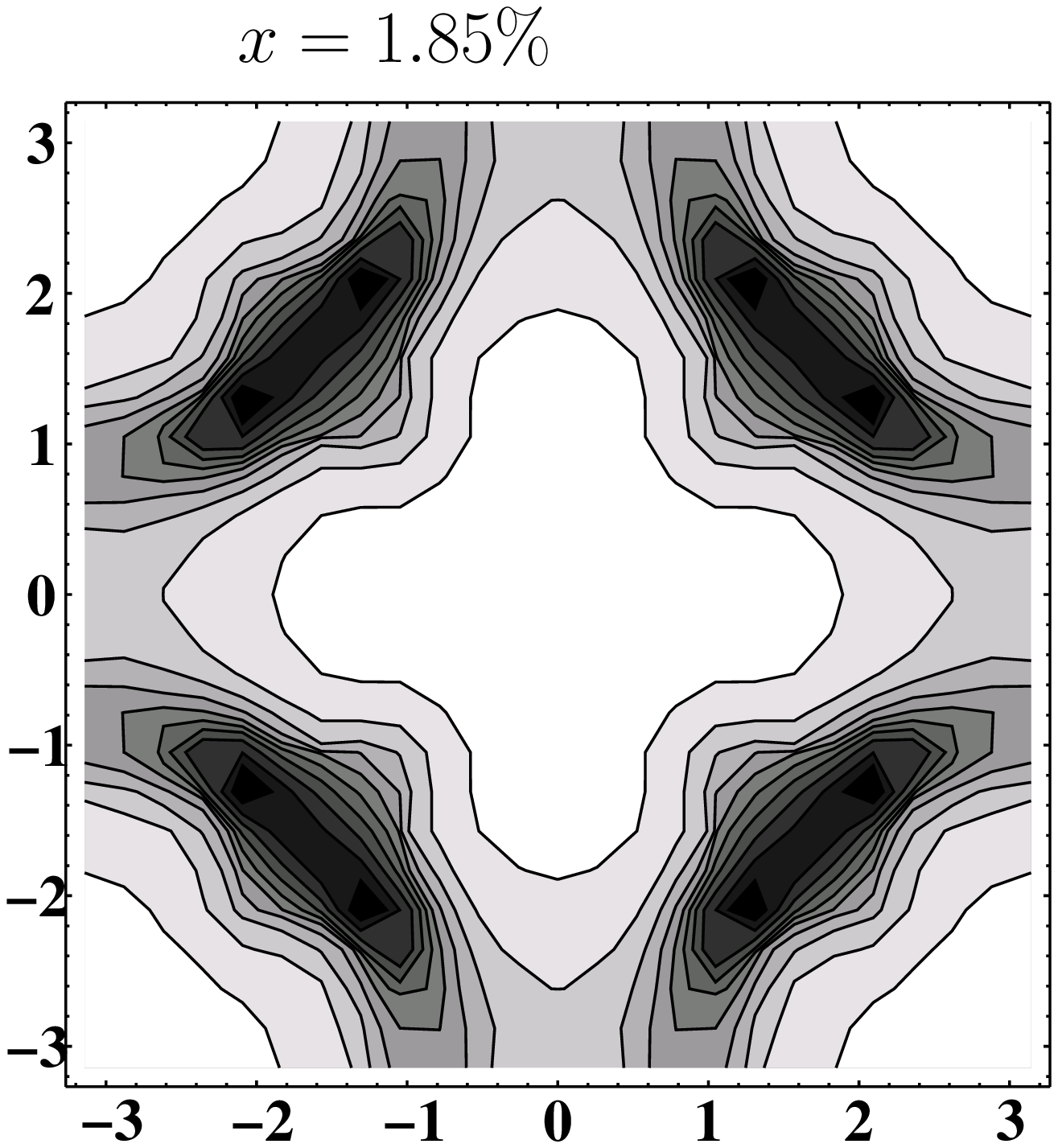} 
\end{center}
\caption{\label{fse}
Low-energy integrated spectrum of the 3BH model for two different electron dopings.
}
\end{figure*}

In order to analyze the evolution of the Fermi surface (FS) as a function
of doping, we plot in Fig.~\ref{fsh} and Fig.~\ref{fse} the low-energy
spectrum in the Brillouin zone for the hole- and electron-doped case, respectively.
This is obtained by integrating the spectrum within an energy window
of width $0.2 t$ around the
Fermi energy.
As discussed above, 
in the hole-doped case, hole pockets start to build around \pit at low
doping, while a large FS centered around $(\pi,\pi)$ develops at larger
doping. This is in qualitative agreement with the experiments 
(see, e.g. Ref. \cite{da.hu.03})
and with
previous results on the 1BH model.
In contrast, in the electron-doped case (Fig.~\ref{fse}), hole pockets
first build around \pio, 
and the FS becomes connected and again closes around 
$(\pi,\pi)$ at larger dopings.

\section{Summary and Conclusions}
\label{summary}

In this paper, we have carried out an analysis of the three-band Hubbard
model in the hole- and in the electron-doped region by means of the
Variational Cluster Approach, whereby we allowed for an
antiferromagnetic, and a superconducting Weiss field, as well as for a
on-site energy as variational parameters for the cluster reference system.
Results for the single-particle spectrum and for the phase diagram
are qualitatively, and to some extent even quantitatively, very
similar to the ones obtained for the one-band Hubbard model with a
next-nearest-neighbor hopping included in order to break particle-hole
symmetry~\cite{se.la.05,ai.ar.05,ai.ar.06}.
Concerning the phase diagram, we obtained a mixed AF+SC phase at low
doping with a transition to a pure SC phase accompanied by phase
separation.
On the basis of these results we can confirm, as already argued in
previous work,  that low-energy
single-particle properties of the three-band Hubbard model
can be quite well captured by a single-band Hubbard model with
appropriate longer-range hopping parameters both in the hole as well
as in the electron-doped case.
Of course, we may expect that two particle excitations, especially
charge-transfer but also spin susceptibilities may behave differently
in the 1BH and the 3BH models. 
Investigations 
along this line are in progress, by means of 
a recently developed extension
of the VCA to treat dynamical two-particle correlation
functions~\cite{br.ar.08u}.

\ack

This work is supported by the Austrian Science Fund (FWF
projects P18551-N16 and J2760-N16),
 by the 
DFG Research Unit n. 538,
and the
NSF grant DMR-0706020.

\bibliographystyle{prsty}
\bibliography{references_database,footnotes}

\end{document}